\documentclass[aps,prr,twocolumn,longbibliography]{revtex4-1}

\usepackage[utf8]{inputenc}
\usepackage{amsmath,amssymb}
\usepackage{amsthm}
\usepackage{graphicx}
\usepackage[colorlinks=true,
  urlcolor=blue,
  linkcolor=blue,
  citecolor=blue]{hyperref}
\usepackage{xcolor}

\usepackage{enumerate}

\newtheorem{theorem}{Theorem}
\renewcommand*{\thetheorem}{\arabic{theorem}}

\newtheorem{lemma}{Lemma}

\newcommand{\ii}{{\rm i}}
\newcommand{\tF}{\tilde{F}}
\newcommand{\Fqr}{F_{q,r}}
\newcommand{\tFqr}{\tilde{F}_{q,r}}

\makeatletter
\renewcommand*{\p@subsection}{}
\makeatletter

\newcommand{\supp}{{Appendix}}
\newcommand{\diff}[1]{{#1}}

\begin{document}

\title{Classification of same-gate quantum circuits and their space-time symmetries with application to the level-spacing distribution}
\author{Urban Duh and Marko Žnidarič}
\affiliation{Physics Department, Faculty of Mathematics and Physics, University of Ljubljana, 1000 Ljubljana, Slovenia}

\begin{abstract}
	We study Floquet systems with translationally invariant nearest-neighbor 2-site gates. Depending on the order in which the gates are applied
	on an $N$-site system with periodic boundary conditions, there are
	factorially many different circuit configurations. We prove that there are
	only $N-1$ different spectrally equivalent classes which can be viewed either
	as a generalization of the brick-wall or of the staircase configuration.
	Every class, characterized by two integers, has a nontrivial space-time
	symmetry with important implications for the level-spacing distribution -- a
	standard indicator of quantum chaos. Namely, in order to study chaoticity one
	should not look at eigenphases of the Floquet propagator itself,
	but rather at the spectrum of an appropriate root of the propagator.
\end{abstract}
\maketitle

\section{Introduction}

Chaoticity and integrability are important theoretical notions. Integrability can allow for analytical results, while chaotic systems, in spite of unpredictability of trajectories, adhere to statistical laws.
So-called toy models -- the simplest models with a given property -- play an
important role. Classical single-particle $H$ in one dimension (1D) is always integrable; one needs at least a 3D phase space for chaos to be possible. This can also be achieved already in 1D~\cite{ott} by a time-dependent $H(t)$, the simplest case being a ``kicked'' system of form $H(t)=\frac{p^2}{2}+V(q)\tau \delta_\tau(t)$, where $\delta_\tau(t)$ is a train of delta functions. A canonical example is the standard map~\cite{stdmap}. Similar logic of taking a Floquet propagator (map) works for quantum systems as well. For instance, one can exemplify single-particle quantum chaos with a
kicked top~\cite{haakeQuantumSignaturesChaos2010,haake_classical_1987}. Going
to many-body quantum systems a plethora of possibilities opens up, one choice being a Floquet propagator that is a product of two simpler propagators, e.g.
Ref.~\cite{Prosen99}. In light of experimental advances in noisy quantum simulations~\cite{Bloch,google,ibm} it pays off to consider systems where the basic building block is a nearest-neighbor gate rather than the local Hamiltonian (applying two-site gates is also simpler in classical numerical simulations~\cite{ulrich}).

We therefore focus on circuits where a one-unit-of-time Floquet propagator $F$
is composed of applying {\em the same} two-site unitary gate $V$ on all
nearest-neighbor pairs of qubits in 1D -- we call such systems {\em simple circuits}. Simple circuits have translational and temporal invariance, and,
depending on the chosen gate $V$, span all dynamical regimes from
integrability to chaos. Needles to say, such simple circuits have been extensively studied, a non-exhaustive list of only few of recent papers includes
Refs.~\cite{Guhr,Gritsev,XXZ,Sarang,Tianci,Katja,Bruno,Chan22,Adam,Pieter,MZ}. We show that any such circuit has a very simple form: it is a product of a local 2-site transformation and a free propagation (translation), and that there are only $N-1$ spectrally inequivalent classes.

There are many indicators of quantum chaos with perhaps the most frequently
used one being the so-called level spacing distribution (LSD) $P(s)$ of
nearest-neighbor eigenenergy spacing
$s$~\cite{haakeQuantumSignaturesChaos2010}. According to the quantum chaos
conjecture~\cite{bohigasCharacterizationChaoticQuantum1984} Hamiltonian systems
with a chaotic classical limit are expected to display $P(s)$ given by
  the random matrix theory (RMT)~\cite{Mehta}. \diff{RMT LSD has also been
  observed in non-integrable generic systems without a classical limit, of
  which simple circuits are an example, where it
is sometimes even taken as a defining property of quantum chaos~\footnote{For many-body quantum systems a word
	of caution is in place. Because the level spacing is exponentially small in system size $N$
	the LSD probes properties on an exponentially large timescale. There can be
	situations where the relevant physics is not chaotic at all despite the RMT
	LSD, see e.g. Refs.~\cite{Lea,Marlon}}.}
For Floquet systems
checking for quantum chaos via $P(s)$ is even simpler: writing eigenvalues of
$F$ as ${\rm e}^{\ii \phi_j}$ the density of eigenphases $\phi_j$ should be
uniform and therefore taking for $s=(\phi_{j+1}-\phi_j){\cal N}/2\pi$,
\diff{where $\mathcal{N}$ is the Hilbert space dimension}, 
there is no need for the unfolding that is required when
studying spectra of $H$~\cite{haakeQuantumSignaturesChaos2010}. It is therefore
rather surprising that, while there are hundreds of papers using $P(s)$ to
study chaoticity in many-body Hamiltonian systems, there are essentially none
studying LSD in simple (same-gate) quantum circuits (exceptions are recent Refs.~\cite{master,huang2023outoftimeorder}). The reason is that, surprisingly,
LSD for chaotic simple circuits seemingly does not adhere to the RMT
expectation. In our paper we will show that the reason behind it is a space-time
symmetry that all such circuits posses.

\begin{figure}[ht]
	\centering
	\includegraphics[width=0.49\textwidth]{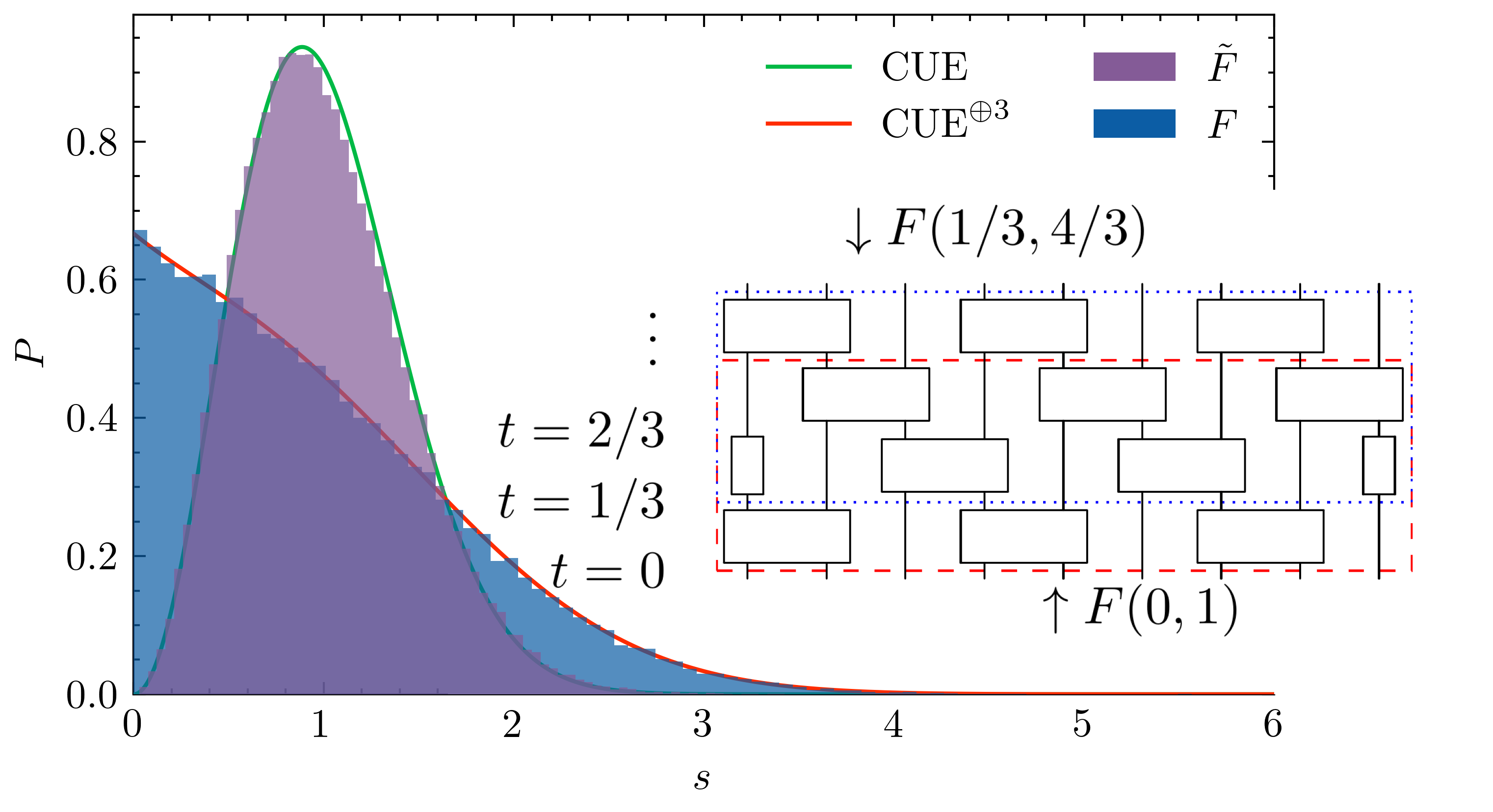}
	\caption{A chaotic 3-layer BW circuit (inset) and the eigenphases
		level-spacing distribution $P(s)$. Eigenphases of the propagator $F$ (blue)
		do not follow the RMT expectation, while after resolving the space-time
		symmetry the eigenphases of the root $\tF=(S^6 F)^{1/3}$ (purple) do agree
		with the CUE RMT (green curve). Red curve is the theory for a direct sum of 3
		CUE matrices (see Eq.~(\ref{eq:rains}) and \supp~\ref{app:numerical}). Data is for
		$N=12$ in the eigenspace with momentum $0$, and Haar random gate $V$.}
	\label{fig:n32_both}
\end{figure}

Let us demonstrate that by a simple generalized brick-wall (BW) circuit
with 3 layers (Fig.~\ref{fig:n32_both} inset), where we translate each 2-site
gate by 3 sites (instead of 2 as in BW). For periodic boundary conditions and
$N$ divisible by $3$ the Floquet propagator can be written as $F = f_2 f_3
	f_1$, where $f_j = \prod_{k = 0}^{N/3 - 1} V_{j + 3k, j + 3k+1}$ is one layer
beginning at site $j$ and $V_{i, j}$ denotes the unitary 2-site gate $V$ acting
on qubits $i$ and $j$. Indices are taken modulo $N$, with sites $j = 1, \dots,
	N$.
Taking a 2-qubit gate $V$ to be some fixed generic unitary, and therefore
having a system that should be quantum chaotic, we can see in
Fig.~\ref{fig:n32_both} that, after resolving the obvious translational symmetry,
the LSD of $F$ is far from the expected RMT result
for a circular unitary ensemble (CUE)~\cite{haakeQuantumSignaturesChaos2010}.
If anything, it is closer to a Poisson statistics typical of integrable
systems, as if there would be some unresolved symmetry~\cite{Olivier}.

Indeed, each layer $f_j$, and thereby also $F$, is invariant under translation by 3 sites, $S^{-3} f_j S^3 = f_{j}$, where $S$ is the translation operator by 1 site to the left,
\begin{equation}
	V_{i + 1, j + 1} = S^{-1} V_{i, j} S.
	\label{eq:S}
\end{equation}
We can also easily see (Fig.~\ref{fig:n32_both}) that translating $F$ by 2
sites is the same as a shift in time by 1 layer. Denoting propagator from time
$t_1$ to $t_2$ by $F(t_1,t_2)$, e.g., $F=F(0,1)$, the 3-layer BW circuits has a
space-time symmetry $S^{-2} F(0,1) S^2 = F(1/3,4/3)$ (application
of each gate advances time by $1/N$). This symmetry is also reflected in the
structure of $F$, which can be written as $F=S^{-4}f_1 S^4 S^{-2}f_1 S^2 f_1=S^{-6} (S^2 f_1)^3=S^{-6} \tF^3$. We now see where the crux of the problem lies. Since $F$ as well as $f_1$ 
have translational symmetry by 3 sites the momentum
$k$ labeling eigenvalues of $S^3$ (\diff{also referred to as the quasi-momentum,
since it takes a discrete set of values}) is a good quantum number. In each momentum
eigenspace $S^{-6}$ is just an overall phase, and
therefore $F$ is, up-to this irrelevant phase, equal to the 3rd power of
$\tF=S^2 f_1$. The quantum chaos conjecture should therefore be applied to
$\tF$ rather than to $F$. Doing that one recovers perfect agreement with the
RMT (Fig.~\ref{fig:n32_both}). It also tells us that, provided $\tF$ and
therefore the circuit is ``chaotic'', the LSD of $F$ will be equal to that of
the 3rd power of a CUE matrix which is equal to a direct sum of 3 independent CUE matrices, Eq.~\eqref{eq:rains}.

The above example is just one possible simple circuit -- in our classification
it is of type $(q,r)=(3,2)$. We shall classify symmetries of all possible simple
quantum circuits, showing that all posses an appropriate space-time symmetry.
Space-time symmetries have been discussed before in the solid-state physics
context and time-periodic
$H(t)$~\cite{xuSpaceTimeCrystalSpaceTime2018,morimotoFloquetTopologicalPhases2017,peng_floquet_2019}.
An important offshoot will be expressing $F$ essentially as a power of simpler matrix $\tF$, meaning that in order to probe dynamic (chaotic)
properties one needs to study $\tF$ and not $F$. Expressions
of that form have appeared before for special cases of the BW
circuit in Ref.~\cite{saIntegrableNonunitaryOpen2021} (class $(2,1)$), and for $r=1$ in Ref.~\cite{huang2023outoftimeorder}, see also
Ref.~\cite{master} for preliminary results.

\section{Classification of simple circuits}

Having an $N$-site 1D system with
periodic boundary conditions there are $N$ nearest-neighbor gates $V_{j,j+1}$
that can be ordered in $N!$ different ways (configurations) to make a one-step
propagator $F$~\footnote{Single-site Hilbert space can have any dimension
	(qudits) although we will frequently speak about qubits because all our
	numerical examples will be of that kind.}. However, it is clear that many of
those configurations have equal $F$ since 2-site gates acting on non-overlapping
nearest neighbor sites commute. Furthermore, a lot of configurations lead to the
same spectra, for instance, under cyclic permutations of gates spetra do not change~\cite{bensaFastestLocalEntanglement2021}. Because we want to study spectra of $F$ we will call two circuits {\em equivalent} if they have the same spectrum. It is clear that there are much less than $N!$ non-equivalent simple circuit classes. For open boundary conditions there is in fact just one class~\cite{bensaFastestLocalEntanglement2021}.

\begingroup
\squeezetable
\begin{table}[t!]
	\begin{ruledtabular}
		\begin{tabular}{rcccc}
			             & \phantom{abc} & \multicolumn{3}{c}{{\bf Allowed $(q,r)$}}                                    \\
			%\cmidrule(r){3-5}
			$\mathbf{N}$ &               & generalized S                             &  & generalized BW                \\
			% \midrule  
			\noalign{\vskip 3pt} \hline \noalign{\vskip 3pt}
			6            &               & (6,1),(6,5)                               &  & (2,1),(3,1),(3,2)             \\
			7            &               & (7,1),(7,2),(7,3),(7,4),(7,5),(7,6)       &  &                               \\
			8            &               & (8,1),(8,3),(8,5),(8,7)                   &  & (2,1),(4,1),(4,3)             \\
			9            &               & (9,1),(9,2),(9,4),(9,5),(9,7),(9,8)       &  & (3,1),(3,2)                   \\
			10           &               & (10,1),(10,3),(10,7),(10,9)               &  & (2,1),(5,1),(5,2),(5,3),(5,4) \\
		\end{tabular}
	\end{ruledtabular}
	\caption{List of $(N-1)$ allowed shift parameters $(q,r)$ for few small $N$, classifying all possible spectrally equivalent $N$ qubit circuits with periodic boundary conditions.}
	\label{tab:qr}
\end{table}
\endgroup
\begin{figure*}[ht!]
	\centering
	\includegraphics{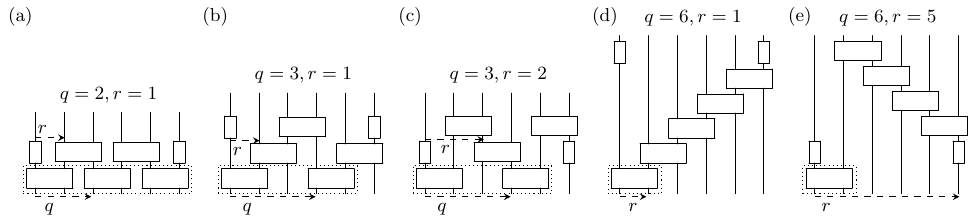}
  \caption{Any same-gate nearest-neighbor circuit is spectrally equivalent to
  one of the canonical circuits $\Fqr$ shown here for $N=6$, see also
Table~\ref{tab:qr}. \diff{Dashed rectangles denote $f_1$, one layer occurring in the root $\tFqr$ in Eq.~\eqref{eq:tFqr}. Circuits (a), (b), and (c) are generalized BW while (d) and (e)
are generalized S. Circuits (b) and (c), as well as (d) and (e), are chiral pairs.}}
	\label{fig:fqr_diagram}
\end{figure*}
For periodic boundary conditions this is not the case. One can show (see \supp,
Theorem \ref{th:equivalence_n}) that there are
$(N-1)$ different equivalence classes. A canonical representative
circuit of a given equivalence class follows a similar logic as the 3-layer BW
example in the introduction: a canonical circuit is characterized by two integers
$q$ and $r$, where the first layer of gates $f_1$ is made by repeatedly
translating $V_{1,2}$ by $q$ sites ($q=3$ in the example), whereas $r$
determines the shift of the 2nd layer with respect to the 1st one ($r=2$ in the
example). The following theorem embodies the precise statement.
\begin{theorem}
	\label{th:equivalence_qr}
	Any simple qubit circuit on $N$ sites with periodic boundary conditions is equivalent to exactly one of the $N - 1$ canonical simple qubit circuits having Floquet propagator
	\begin{equation}
		\Fqr = S^{-qr} \left(S^r (S^q V_{1, 2})^{N/q}\right)^{q}, \label{eq:qr_def}
	\end{equation}
	where $q$ is larger than $1$ and divides $N$, and $q$ and $r$ are coprime,
	\begin{align}
		 &  & 2 \leq q \leq N, &  & \mathrm{gcd}(q,N)=q,\nonumber                \\
		 &  & 1 \leq r<q,      &  & \mathrm{gcd}(q, r) = 1, \label{eq:qrn_cases}
	\end{align}
	where $\mathrm{gcd}$ denotes the greatest common divisor.
\end{theorem}
The proof can be found in \supp~\ref{app:proof}. While it is not constructive
in the sense of providing an explicit procedure of transforming a given $F$ to
its canonical form $\Fqr$, the transformation is in practice easily achieved by
hand for small $N$, or one can in linear time
calculate the invariant $p$ introduced in Lemma~\ref{lm:c_conserv} in
\supp~\ref{app:proof}, thereby obtaining the correct $(q,r)$. The integer
invariant $p$ is defined for an alternative simple circuit representative of a
given class, and characterizes the circuit as a concatenation of two staircase sections with opposite chirality (Fig.~\ref{fig:fnn_diagram}), the length of the 2nd being $p$.

The canonical form of $\Fqr$ in Eq.~\eqref{eq:qr_def} has a simple geometric
interpretation: the term in the inner bracket is a single layer $f_1$ that is
composed of $N/q$ gates, which is then with appropriate shifts repeated in
altogether $q$ layers, explicitly written as
\begin{equation}
	\Fqr = \prod_{j = 0}^{q - 1} \prod_{i = 0}^{N/q - 1} V_{1 + jr + iq, 2 + jr + iq}.
\end{equation}
The $(N-1)$ classes described in Theorem~\ref{th:equivalence_qr} account for
all possible circuits of which the standard brick-wall with $(2,1)$, and the
staircase~\cite{Sagar,Green,bensaFastestLocalEntanglement2021} (also called
convolutional codes~\cite{Poulin,Hayden}) with $(N,1)$ are just two cases. The
allowed values of $(q,r)$ for few small $N$ are listed in Table~\ref{tab:qr},
while their pictures are shown for $N=6$ in Fig.~\ref{fig:fqr_diagram},
and for $N=10$ in Fig.~\ref{fig:fqr_diagram_10} in \supp. Note that while the allowed set of
$(q,r)$ for a given $N$ depends on the factors of $N$, the allowed $N$s for a
given $(q,r)$ are simpler: taking any coprime $q>r$ a circuit is possible
for all $N$ that are multiples of $q$.

The set of allowed $(q,r)$ naturally splits into two categories.
Because $q$ divides $N$, with the maximal value being $N$, one group is
composed of the largest possible $q=N$, while the other has smaller $2 \le q
	\le N/2$. Group (i) are {\em generalized S} circuits with $q=N$. Because the
translation by $N$ is equivalent to the identity this group could be
equivalently described by $(N,r)\equiv (r,0)$, that is, by a single integer $r$
that gives the shift of the next gate. As $r$ is coprime with $N$ all n.n.\
gates are obtained by \diff{just this} translation modulo $N$. Group (ii) can
be viewed as {\em generalized BW} circuits and needs two integers. Because $q$
divides $N$, translation by $q$ alone \diff{does not generate all n.n\ gates} 
and one needs subsequent layers characterized by $r$. 
Altogether one has a $q$-layer BW circuit, each layer consisting of $N/q$ gates.

\begin{figure*}[t!]
	\centering
	\includegraphics[width=\textwidth]{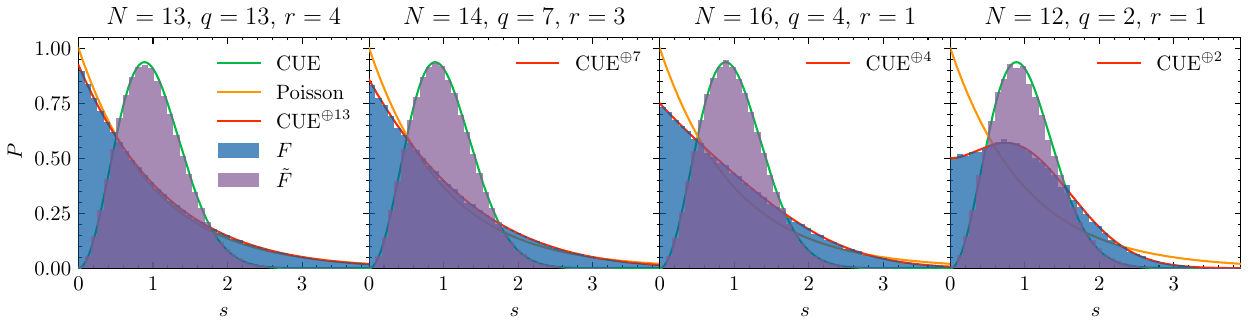}
	\caption{Level-spacing distribution $P(s)$ of \diff{eigenphases of} $\Fqr$ (Eq.~\ref{eq:qr_def}) in blue are not chaotic and are equal to a direct sum of $q$ CUE matrices (red curve, \supp~\ref{app:numerical}), while the LSD of the \diff{eigenphases of} root $\tFqr$ (Eq.~\ref{eq:tFqr}, violet) agrees with the CUE RMT prediction (green curve). For $q \neq N$ we show data from the momentum eigenspace of $S^q$ with $k = 0$, and averaging is done over $10-50$ circuits \diff{each having an independent 2-site Haar random gate $V$}.}
	\label{fig:level_spacings}
\end{figure*}
For a generic gate $V$ the spectra of all $(N-1)$ propagators $\Fqr$ are
different. If $V$ would be symmetric with respect to the exchange of the
two qubits the circuits with $(q,r)$ and $(q,q-r)$ would have the same spectra
(spatial reflection symmetry), and therefore one would have only $\lfloor
	\frac{N}{2} \rfloor$ different spectral
classes~\cite{bensaFastestLocalEntanglement2021}.
Theorem~\ref{th:equivalence_qr} also shows that for prime $N$ only
generalized S circuits exist. For odd $N$ there are no standard BW circuits having
$(2,1)$, but there are generalized BW circuits with $q>2$ (see Table
\ref{tab:qr}). \diff{It is interesting to note that circuits with more complex multi-site update rules have been used before, for instance the $(3,1)$ case~\cite{Balazs21} as well as $(4,1)$ has been used to construct integrable models~\cite{Balazs21PRE} (although with a 3-site transformation). One interesting question is possible integrability of different canonical configurations for specific $V$.} While for smaller $N<10$ possible circuits are
straightforward generalizations of the S or BW configurations with left or
right chirality, for larger $N$ less intuitive circuits are also possible. For
instance, for $N=10$ one can have $(5,2)$ (see Fig.~\ref{fig:fqr_diagram_10} in \supp) that can be further compressed in time direction (e.g., all gates in the first
two layers $f_1$ and $f_3$ commute), reducing the number of non-commuting
layers from $q=5$ to just $3$. Each of those compressed layers has $2$
idle qubits (that are not acted upon), such that the compressed circuit
$F_{5,2}^t$ has two separate \diff{diagonally slanted} lines of idle qubits,
each of width $1$. Integers $(q,r)$ therefore also determine the filling
fraction, i.e., the number and the pattern of idle qubits in maximally compressed
$\Fqr^t$ (see \supp~\ref{app:diagrams}). The only circuit with no
idle qubits is the standard BW with $(2,1)$.

\section{Space-time symmetries}

In order to understand space-time symmetries of any simple circuit it suffices to study the canonical equivalence class representatives $\Fqr$. Denoting the inner term in Eq.~(\ref{eq:qr_def}) by $\tFqr$, calling it a root of $\Fqr$,
\begin{align}
	\tFqr & = S^r (S^q V_{1, 2})^{N/q} = S^r f_1,
	\label{eq:tFqr}
\end{align}
where $f_1=\prod_{k = 0}^{N/q - 1} S^{-kq} V_{1, 2} S^{kq}$, we can write Eq.~(\ref{eq:qr_def}) as
\begin{align}
	\Fqr = S^{-qr} (\tFqr)^q. \label{eq:f_pow_q}
\end{align}
Eq.~(\ref{eq:f_pow_q}) appeared in Ref.~\cite{saIntegrableNonunitaryOpen2021} in the open-systems context for the special case of a BW circuit with $(2,1)$. The root connection is especially simple for the generalized S case: for $q=N$ the translation $S^q$ is identity, resulting in $\tFqr = S^r V_{1, 2}$ and $\Fqr = (S^r V_{1, 2})^N$.

The root has a translational symmetry by $q$ sites
\begin{align}
	S^{-q} & \tFqr S^q = \tFqr.
\end{align}
This is trivially true if $q = N$, otherwise the translated root is equal to
$S^r \prod_{k = 0}^{N/q - 1} S^{-(k + 1)q} V_{1, 2} S^{(k + 1)q}$, where, since
all the gates in the product commute, we can relabel the index $k +1=k'$, yielding $S^r \prod_{k' = 0}^{N/q - 1} S^{-k' q} V_{1, 2} S^{k' q}=\tFqr$. Because $\Fqr$ is a power of $\tFqr$ multiplied by some power of $S^q$, $\Fqr$ also has translational symmetry by $q$ sites.

Furthermore, $\Fqr$ also has a space-time symmetry when translating by $r$ sites
\begin{align}
	S^{-r} \Fqr(0,1) S^r = \Fqr(1/q,1 + 1/q). \label{eq:st_sym_div}
\end{align}
This can be easily seen by rewriting the LHS of Eq.~\eqref{eq:st_sym_div}
as $S^{-qr} \left(S^{r} S^{-r} (S^q V_{1, 2})^{N/q} S^r\right)^q$, which then equals
to $S^{-qr} \left(S^{r} (S^q V_{1 + r, 2 + r})^{N/q}\right)^q$. The final expression
can be understood as a circuit beginning with the second layer, thus justifying the equality to the RHS of Eq.~\eqref{eq:st_sym_div}. Eq.~(\ref{eq:st_sym_div}) appeared in Ref.~\cite{huang2023outoftimeorder} for the special case of $r=1$, \diff{along with a figure showing numerically computed LSD of $\Fqr$ for $q=2, r=1$}.

\section{Level spacing statistics}

An immediate application of the above results
is in quantum chaos for the statistics of spacings of closest eigenphases of
$F$. Looking at the root connection in Eq.~(\ref{eq:f_pow_q}) and the fact that
$S^q$ commutes with all terms, one can focus on a given common momentum
eigenspace of $S^q$ with eigenvalues $e^{2\pi\ii k q/N}$, $k \in \{0, 1, \dots, N/q - 1\}$.
There $S^{-q r}$ is just an overall phase factor $e^{-2\pi\ii q k r/N}$.
Therefore $\Fqr$ is up-to this phase equal to an appropriate power of $\tFqr$.

\diff{The eigenphases of $\Fqr$ are therefore simple $q$-th multiples of eigenphases of $\tFqr$ modulo $2\pi$ (and adding the momentum phase factor). For high $q$ such an operation will results in an uncorrelated Poisson statistics of eigenphases of $\Fqr$\footnote{Provided the eigenphases $\tilde{\phi}_j$ of $M$ are incommensurate with $2\pi$ this will always hold for large $q$ regardless of the properties of $M$ because the map $\phi_j = q\cdot \tilde{\phi}_j\, ({\rm mod}\, 2\pi)$ is ergodic \diff{(sawtooth map)} and will destroy any correlations in $\tilde{\phi}_j$.}, i.e., an exponential distribution of $P(s)$, and to infer possible quantum chaos one should not look at the eigenphases of $\Fqr$. Rather, if the circuit is quantum chaotic one would expect that the spectral
statistics of the root $\tFqr$ (\ref{eq:tFqr}) will adhere to the RMT theory.} In particular, if the 2-site gate $V$ does not have any anti-unitary
(time-reversal) symmetry, which is the case for our numerics where $V$ is
randomly picked according to the Haar measure, the appropriate ensemble \diff{for $\tFqr$ is the
CUE (i.e., the unitary Haar measure)}. We can see in Fig.~\ref{fig:level_spacings} that the LSD of $\tFqr$ indeed
agrees with CUE Wigner surmise $P(s) = \frac{32}{\pi^2} s^2 e^{-s^2 4/ \pi}$ for all canonical
classes.

\diff{If one is on the other hand interested in the eigenphases of $\Fqr$ one has to take into account the nontrivial modulo $2\pi$ operation.} The theorem by
Rains~\cite{rainsImagesEigenvalueDistributions2003} tells us what is the
distribution \diff{of eigenvalues} of a power of a matrix \diff{from the unitary Haar measure. For $M
\in \mathrm{U}_{\mathcal{N}}$, where
$\mathrm{U}_{\mathcal{N}}$ denotes the Haar distribution of $\mathcal{N} \times
\mathcal{N}$ unitary matrices, the eigenvalues of its power $M^q$ are a union of $q$ independent eigenvalue sets distributed according to $\mathrm{U}_{{\cal N}_j}$ of smaller matrices,}
\begin{align}
	\diff{\lambda(M^q) \sim \bigcup_{j=0}^{q-1}\, \lambda\left( \mathrm{U}_{{\cal N}_j} \right),\quad {\cal N}_j=\left\lceil \frac{\mathcal{N} - j}{q} \right\rceil,} \label{eq:rains}
\end{align}
\diff{for any $q<\mathcal{N}$, $\lceil \bullet \rceil$ denotes the ceiling function, and $\lambda(\mathrm{U}_{\cal N})$ is the distribution of eigenvalues of Haar-random unitary matrices of size ${\cal N}$. Note that ${\cal N}=\sum_{j=0}^{q-1}{\cal N}_j $, and therefore, as far as the eigenvalue distribution is concerned, it is as if $M^q$ would have a block diagonal structure with blocks of smaller CUE matrices. In our case of large $\mathcal{N}$ and small $q \ll {\cal N}$ all dimensions in the union are approximately equal to ${\cal N}_j \approx {\cal N}/q$, which means
that the level-spacing distribution in each eigenspace of $S^q$ of size ${\cal N} \approx 2^Nq/N$} of $\Fqr$
behaves in the same way as if it had another unitary symmetry with $q$ distinct
sectors~\cite{Olivier}. \diff{Theoretical LSD in such a case of a sum of $q$ independent spectra is known and has been studied long time ago~\cite{rosenzweigRepulsionEnergyLevels1960}, see also \supp~\ref{app:numerical} (or Appendix A in Ref.~\cite{Mehta}).} We can see in Fig.~\ref{fig:level_spacings} that this theory agrees with numerical LSD for $\Fqr$ \footnote{For small $N$ one of course has non-negligible finite-size deviations from RMT distributions; we have checked that they are within expectations~\cite{prosenEnergyLevelStatistics1993}.}.

\section{Discussion}
We have classified all different quantum circuits in one
dimension with periodic boundary conditions and translationally invariant nearest-neighbor 2-site gates. There are $(N-1)$ different spectral
classes, being generalizations of the familiar brick-wall and the staircase
configurations. Each class can be characterized by two integers $(q,r)$, such that the Floquet propagator is essentially a $q$-th power of $\tF=S^r f_1$, where for generalized S circuits one has $f_1=V_{1,2}$, while for generalized BW circuits $f_1$ is one layer of gates. We have therefore come full circle: similarly as in classical single-particle kicked models where one interchangeably applies a simple map in real space (e.g., potential $V(q)$) and a simple map in momentum space (e.g., free evolution), any quantum many-body translationally invariant Floquet system has the same basic structure. The elementary building block $\tF$ is a product of simple local transformation, like $V_{1,2}$, and of ``free evolution'' described by the translation operator $S$ (that is diagonal in the Fourier basis).

We have explicitly shown how that affects the level spacing statistics of simple circuit Floquet systems -- to detect quantum chaos one must look at the spectrum of the $q$-th root $\tF$ of the propagator. Effects of the underlying space-time symmetry on other quantifiers of quantum chaos remain to be explored. Reducing all circuits to just few canonical classes makes it possible to study chaoticity for different $(q,r)$; are some configurations more chaotic than others, does that depend on the filling fraction (see \supp~\ref{app:diagrams})? While we focused on systems without any symmetry, i.e., the unitary case, orthogonal and symplectic cases can be treated along the same lines. Generalizing classification to more than one dimension is also an open problem.

\begin{acknowledgements}
  We acknowledge support by Grants No.~J1-4385 and No.~P1-0402 from the Slovenian Research Agency (ARIS).
  \end{acknowledgements}

\bibliography{references}

\clearpage
\appendix

\setcounter{secnumdepth}{2}
\renewcommand{\thesubsection}{\Alph{section}.\arabic{subsection}}

% Different counting
%\renewcommand*{\thetheorem}{S\arabic{theorem}}
%\renewcommand*{\thelemma}{S\arabic{lemma}}
%\renewcommand*{\thefigure}{S\arabic{figure}}
%\setcounter{figure}{0}

\onecolumngrid

\begin{figure*}[ht]
	\centering
	\includegraphics{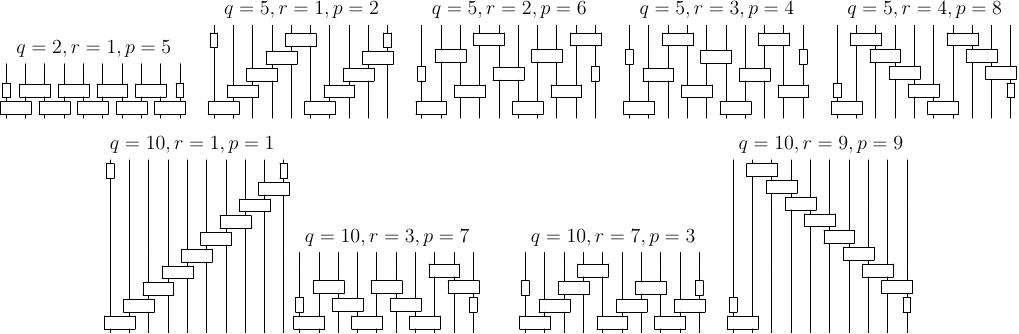}
	\caption{All allowed $F_{q, r}$ defined in Eq.~\eqref{eq:qr_def} for $N = 10$.
		The parameter $p$ of the equivalent $F_p$ (Eq.~\eqref{eq:fnn_def}) is also included. Here
		$C(F_{q, r}) = p$, see discussion in Section~\ref{app_sub:invariant}.}
	\label{fig:fqr_diagram_10}
\end{figure*}

\twocolumngrid

\section{Canonical circuits}
\label{app:diagrams}

In the main text in Fig.~\ref{fig:fqr_diagram} we have shown canonical circuits $\Fqr$ for $N=6$. Here we show in Fig.~\ref{fig:fqr_diagram_10} all $9$ canonical $\Fqr$ for $N=10$, where a variety of configurations is richer. We can for instance notice that even though circuits are constructed as a $q$-layered circuit in some cases consecutive layers commute and can therefore be compressed, thus reducing the number of layers. For $N=10$ this is the case for $(q,r)=(5,2)$ and its chiral pair $(5,3)$. For $(5,2)$ the first two layers $f_1$ and $f_3$ commute and can be compressed to a single layer; likewise for the next two layers. Therefore in the compressed form $(F_{5,2})^2$ consists of only 5 layers instead of $10$. In this compressed form there are still $2$ idle qubits in each layer on which no gate acts. One can say that the filling fraction of gates for the circuit $(5,2)$ is $8/10$, i.e., $20\%$ of qubits is idle. The class $(5,1)$ on the other hand can not be compressed any further and has the filling fraction of only $4/10$. The only circuit with filling fraction $1$ is the standard BW with $(2,1)$.

\section{Level-spacing distribution of a direct sum of independent RMT matrices}
\label{app:numerical}

The general formula for the LSD of a direct sum of independent RMT matrices of
arbitrary dimensions was derived in
Ref.~\cite{rosenzweigRepulsionEnergyLevels1960}. The matrices in the union
in the theorem by Rains~\cite{rainsImagesEigenvalueDistributions2003}, Eq.~\eqref{eq:rains}, are approximately of equal
dimensions in the large $N$ limit, which is why we use the formula for
the direct sum of equally dimensional matrices throughout this paper.

For the LSD $P(s)$ of an RMT ensemble we define
\begin{align}
	R(y) & = \int_0^\infty P(x + y) \, \mathrm{d}x,   \\
	D(y) & = \int_0^\infty x P(x + y) \, \mathrm{d}x,
\end{align}
where $R(y)$ is nothing but $1$ minus the cumulative distribution of $P$.
The LSD for a direct sum of $m$ independent equally dimensional
matrices from the same RMT ensemble is then
\begin{align}
	P_m(s) = D^m(s/m) \left[\frac{1}{m} \frac{P(s/m)}{D(s/m)} + \left(1 - \frac{1}{m}\right) \frac{R^2(s/m)}{D^2(s/m)}\right].
\end{align}
In Figures in the paper, we plot $P_m(s)$ obtained by using the Wigner surmise
for $P(s)$, which is of acceptable accuracy for our application. For better
approximations of $P(s)$ see Ref.~\cite{dietzTaylorPadeAnalysis1990}.

\section{Proof of Theorem 1 from the main text}
\label{app:proof}

In this section, we set to prove Theorem~\ref{th:equivalence_qr}, the main
result of this paper. The proof is divided into subsections containing more
theorems and lemmas. First, we introduce another canonical configuration in
Section~\ref{app_sub:double_staircase}, which can be geometrically interpreted
as a concatenation of two staircase circuits with opposite chiralities. It is
useful for determining that the number of non-equivalent circuits is $(N - 1)$
and later used in the proof of Theorem~\ref{th:equivalence_qr}. After that in
Section~\ref{app_sub:invariant}, we introduce the quantity $C(F)$ invariant for
equivalent circuits and state its important properties. This invariant is
crucial in later proofs but is also useful on its own to efficiently determine
the canonical form of a given circuit. Finally, we focus on the generalized
S/BW circuits $\Fqr$ in Section~\ref{app_sub:fqr}, where we combine the results
from previous sections to prove Theorem~\ref{th:equivalence_qr}.

To keep our notation clearer, we identify a product of $N$ 2-site gates acting on
nearest neighbor sites with a sequence of $N$ numbers in the following way
\begin{equation}
	V_{i_N, i_N + 1} \cdots V_{i_2, i_2 + 1} V_{i_1, i_1 + 1} \equiv (i_1, i_2, \dots, i_N).
\end{equation}
We will refer to $i_l$ as gate numbers and $l$ as time indices. When talking about
the gates appearing before/after a gate $i_l$ in a given $F$, we will call $i_{l + 1}$
the time successor and $i_{l - 1}$ the time predecessor, whereas when referring
to gates with neighboring numbers, we will call $i_l - 1$ the left neighbor and
$i_l + 1$ the right neighbor of $i_l$.
By definition, the sequences corresponding to simple circuits are permutations of the
first $N$ natural numbers. For the canonical simple circuits defined in the paper
in Eq.~\ref{eq:qr_def}
\begin{align}
	\Fqr \equiv ( & 1, 1 + q, \dots, 1 + (N/q - 1)q, \nonumber   \\
	              & ,1 + r, 1 + r + q, \dots), \label{eq:qrn_seq}
\end{align}
where all gate numbers are taken modulo $N$ (from $1$ to $N$).

We are interested in (spectrally) equivalent circuits, as defined in the paper. The
equivalence will be denoted with $\cong$. The two important
equivalence operations are time predecessor/successor commutation of non-neighboring gates (here
the Floquet operators are actually equal, not only equivalent)
\begin{align}
	|i_k - i_{k + 1}| & > 1 \pmod{N} \quad \implies \nonumber                                                                                                \\
	V_{i_N, i_N + 1}  & \cdots V_{i_{k + 1}, i_{k + 1} + 1} V_{i_k, i_k + 1}  \cdots V_{i_1, i_1 + 1} =                                 \label{eq:com_equiv} \\
	                  & = V_{i_N, i_N + 1} \cdots V_{i_k, i_k + 1} V_{i_{k + 1}, i_{k + 1} + 1}\cdots V_{i_1, i_1 + 1}, \nonumber
\end{align}
in the new notation written as
\begin{align}
	 & |i_k - i_{k + 1}| > 1 \pmod{N} \quad \implies \label{eq:nn_comm_seq}                                 \\
	 & \ (i_1, \dots, i_k, i_{k + 1}, \dots, i_N) \cong (i_1, \dots, i_{k + 1}, i_k, \dots, i_N), \nonumber
\end{align}
and cyclic permutation
\begin{align}
	V_{i_N, i_N + 1} \cdots & V_{i_2, i_2 + 1} V_{i_1, i_1 + 1} \cong \nonumber                                      \\
	                        & \cong V_{i_1, i_1 + 1} V_{i_N, i_N + 1} \cdots V_{i_2, i_2 + 1}, \label{eq:circ_equiv}
\end{align}
or in the new notation
\begin{align}
	 & (i_1, i_2, \dots i_N) \cong (i_2, \dots, i_N, i_1). \label{eq:cyclic_perm_seq}
\end{align}

While the equivalence of cyclically permuted circuits was motivated by spectral
equivalence, a different, perhaps more general, motivation is also possible. If
we have dynamics in mind, i.e., $F^t$ with possibly large $t$, the definition
of a starting time of our period is arbitrary, e.g., the first operator in a
period could just as well have been defined as the last operator in the
previous period. Therefore, it would also make sense to define cyclically
permuted circuits to be equivalent in this case.

\subsection{Double staircase canonical circuits $F_p$}
\label{app_sub:double_staircase}

\begin{figure}[h]
	\centering
	\includegraphics{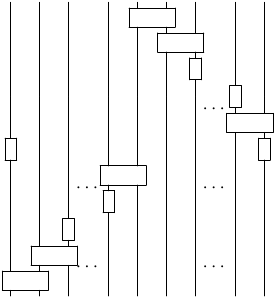}
	\caption{Diagram of $F_p$ defined in Eq.~\eqref{eq:fnn_def}.}
	\label{fig:fnn_diagram}
\end{figure}

We now introduce a new canonical form, different from the one in the main paper, its diagram
is shown in Fig.~\ref{fig:fnn_diagram}. It consists of two staircase sections
with opposite chirality, the second having length $p$. To show that it is
indeed a canonical form, i.\ e.\ that every simple circuits is equivalent to
some circuit in this form, we can prove the following Theorem.
\begin{theorem}
	\label{th:equivalence_n}
	Any simple  circuit is equivalent to a simple circuit with the Floquet operator
	given by
	\begin{align}
		F_p = \   & V_{N - p + 1, N - p + 2} \cdots V_{N - 1, N} V_{N, 1} \nonumber  \\
		          & V_{N - p, N - p + 1} \cdots V_{2, 3} V_{1, 2} \label{eq:fnn_def} \\
		\equiv \  & (1, 2, \dots, N - p, N, N - 1, \dots, N - p + 1), \nonumber
	\end{align}
	for some $p \in \{1, 2, \dots, N - 1\}$.
\end{theorem}
\begin{proof}
	Let $F = (i_1^{(0)}, i_2^{(0)}, \dots, i_N^{(0)})$ be any simple circuit.
	This means that $i_k^{(0)} \neq i_l^{(0)}$ for $k \neq l$. First, we will
	bring gate $1$ to index $1$ (\textit{step 1}). After that, we will try to
	bring gate $2$ to index $2$ (\textit{step 2}). This will either be possible,
	in which case we will continue to try bring gate $k$ to index $k$ in the
	\textit{general step}, or it won't be possible, in which case our gate will be
	equivalent to $F_{N - 1}$.

	\textit{Step 1}: By means of cyclic permutations~\eqref{eq:cyclic_perm_seq}
	we can always transform $F$ into an equivalent simple circuit beginning
	with gate $1$ ($V_{1,2}$ in the standard notation)
	\begin{equation}
		F \cong (1, i^{(1)}_2, \dots, i^{(1)}_N),
	\end{equation}
	where we have appropriately relabeled the indices.

	\textit{Step 2}: We now try to bring gate $2 = i^{(1)}_K, K \in \{2, \dots
		N\}$ to the second position by doing time predecessor/successor
	commutations~\eqref{eq:nn_comm_seq}. Gate $2$ does not commute only with gates $1$ and
	$3$, which means that we must consider two possibilities:
	\begin{enumerate}[(i)]
		\item Gate $3$ appears after gate $2$, $3 = i^{(1)}_L$ for $L > K$: \\
		      By applying~\eqref{eq:nn_comm_seq}, we can bring gate $2$ to index $2$
		      \begin{equation}
			      F \cong (1, 2, i^{(2)}_3, \dots, i^{(2)}_N),
		      \end{equation}
		      where we have again relabelled the indices.
		      We can continue with the \textit{general step}.
		\item Gate $3$ appears before gate $2$, $3 = i^{(1)}_L$ for $L < K$: \\
		      By applying~\eqref{eq:nn_comm_seq} (and relabelling the indices), we
		      can bring gate $2$ to be the time successor of gate $3$
		      \begin{equation}
			      F \cong (1, i'^{(1)}_2, \dots, 3, 2, \dots, i'^{(1)}_N).
		      \end{equation}
		      In the context of equivalence transformations, we can now think of the sequence of gates
		      $(3, 2) = (i'^{(1)}_{K'}, i'^{(1)}_{K' + 1})$ as a gate
		      that does not commute only with gates $1$ and $4$. Again, we have two similar cases:
		      \begin{enumerate}[(a)]
			      \item Gate $4$ appears after the sequence of gates $(3, 2)$, $4 = i'^{(1)}_{L'}$ for $L' > K' + 1$: \\
			            By applying~\eqref{eq:nn_comm_seq}, we can bring gate $3$ to index $1$ and gate $2$ to index $3$
			            \begin{equation}
				            F \cong (3, 1, 2, \dots)
			            \end{equation}
			            and by applying~\eqref{eq:cyclic_perm_seq}, we can cyclically permute gate $3$ to index $N$
			            \begin{equation}
				            F \cong (1, 2, i^{(2)}_3, \dots, i^{(2)}_{N - 1}, 3).
			            \end{equation}
			            We can now continue with the \textit{general step}.
			      \item Gate $4$ appears before $(3, 2)$, $4 = i'^{(1)}_{L'}$ for $L' < K'$: \\
			            By applying~\eqref{eq:nn_comm_seq}
			            \begin{equation}
				            F \cong (1, \dots, 4, 3, 2, \dots).
			            \end{equation}
			            We can again think of $(4, 3, 2)$ as a gate that does not commute only with
			            gates $5$ and $1$ and again consider two cases similar to (a) and (b). By repeating this
			            process, we either get to the point where case (a) arises, and we can continue with the
			            \textit{general step}, or our circuit is equivalent to
			            \begin{equation}
				            F \cong (1, N, N - 1, \dots, 2).
			            \end{equation}
			            Here, the obtained equivalent circuit is thus precisely $F_{N - 1}$.
		      \end{enumerate}
	\end{enumerate}

	\textit{General step}: Let us suppose we have already shown that $F$ is equivalent to
	\begin{equation}
		F \cong (1, 2, \dots, k, i^{(k)}_{k + 1}, \dots, i^{(k)}_N),
	\end{equation}
	where $i^{(k)}_l$ are arbitrary indices relabelled in a convenient way.
	We can treat the sequence of gates $(1, 2, \dots, k)$ as a gate, that does not
	commute only with $N$ and $k + 1$, which means that we can repeat step 2 by
	trying to bring gate $k +1$ to the right of $k$ (index $k +1$). Thus,
	either
	\begin{enumerate}[(i)]
		\item $F \cong (1, \dots, k, k + 1, \dots)$ or
		\item $F \cong (1, \dots, k, N, N - 1, \dots, k + 1) = F_{N-k}$.
	\end{enumerate}
	In (i) we can repeat the \textit{general step} until case (ii) arises, or we
	end up with $F \cong (1, 2, \dots, N) = F_1$.
\end{proof}

Since there are $(N - 1)$ allowed $p$, it is clear that there are (at most) $(N
	- 1)$ non-equivalent simple circuits. The proof given is constructive,
which means that we can use it as an algorithm to convert a given $F$ to some
$F_p$.

\subsection{Circuit invariant}
\label{app_sub:invariant}

An alternative way to determine the equivalent $F_p$ for a given $F$ is to calculate
some quantity, which is invariant in equivalence operations and is different for all
$F_p$. A convenient choice is the length of the second staircase $p$, which
is clearly equal to the number of gates, for which their right neighbor (modulo $N$) appears
in the Floquet operator before them. Let us thus define $C(F)$ to be exactly that
\begin{align}
	C(i_1, \dots, i_N) = | \{ & i_l, l \in \{1, \dots, N\}; \exists i_k, k < l: \nonumber \\
	                          & i_k - i_l \equiv 1 \pmod N\}|, \label{eq:c_def}
\end{align}
where $|\bullet|$ denotes the cardinality of a set. As stated, $C(F_p)$ is clearly equal
to the length of the second staircase
\begin{align}
	C(F_p) = |\{N, N - 1, \dots, N - p + 1\}| = p. \label{eq:c_fp_val}
\end{align}
Let us now show that $C(F)$ is indeed invariant under circuit equivalence operations.
\begin{lemma}
	For simple circuits $C(F)$ is invariant under circuit equivalence operations \eqref{eq:nn_comm_seq} and \eqref{eq:cyclic_perm_seq}.
	\label{lm:c_conserv}
\end{lemma}
\begin{proof}
	Let
	\begin{align}
		\tilde C(i_1, \dots, i_N) = \{ & i_l, l \in \{1, \dots, N\}; \exists i_k, k < l: \nonumber \\
		                               & i_k - i_l \equiv 1 \pmod N\},
		\label{eq:c_tilde_def}
	\end{align}
	be the set of gates the invariant is counting, which means $C(i_1, \dots, i_N) = |\tilde C(i_1, \dots i_N)|$.

	As stated previously, for simple circuits $(i_1, \dots, i_N)$ is just a permutation of $(1, \dots, N)$,
	which means
	\begin{align}
		i_1     & \notin \tilde C(i_1, \dots, i_N), \text{because nothing is before} \ i_1       \\
		i_1 - 1 & \in \tilde C(i_1, \dots, i_N), \text{because} \ i_1 - 1 \in \{i_j\}_{j = 2}^N,
	\end{align}
	where indices are taken modulo $N$.
	But in the equivalent cyclically permuted circuit~\eqref{eq:cyclic_perm_seq}
	\begin{align}
		i_1     & \in \tilde C(i_2, \dots, i_N, i_1), \text{because} \ i_1 + 1 \in \{i_j\}_{j = 2}^N \\
		i_1 - 1 & \notin \tilde C(i_2, \dots, i_N, i_1).
	\end{align}
	The membership of all other gates in $\tilde C$ stays the same in both
	cases, since the position of $i_1$ can change only the membership of $i_1 -
		1$ and $i_1$ in $\tilde C$ and all other relative positions stay the same. Thus
	\begin{equation}
		C(i_1, \dots, i_N) = C(i_2, \dots, i_N, i_1),
	\end{equation}
	which means that $C$ is conserved under cyclical permutations~\eqref{eq:cyclic_perm_seq}.

	A transposition of time successive gates can only change $C$ if their gate number difference is $\pm 1$,
	which does not happen if they commute (Eq.~\eqref{eq:nn_comm_seq}). Therefore, $C$ is conserved there
	\begin{align}
		 & |i_k - i_{k + 1}| > 1 \pmod{N} \quad \implies \nonumber                                  \\
		 & \ C(i_1, \dots, i_k, i_{k + 1}, \dots, i_N) = C(i_1, \dots, i_{k + 1}, i_k, \dots, i_N).
	\end{align}
\end{proof}

An obvious consequence of Theorem~\ref{th:equivalence_n} is the following lemma.
\begin{lemma}
	For simple circuits $C(F) \in \{1, \dots, N - 1\}$.
	\label{lm:c_values}
\end{lemma}
\begin{proof}
	According to Theorem~\ref{th:equivalence_n}, any simple circuit $F$ is
	equivalent to some $F_p$ for $p \in \{1, \dots, N - 1\}$. Since $C(F)$
	is conserved under equivalence transformations (Lemma~\ref{lm:c_conserv}) and $C(F_p) = p$,
	$C(F) \in \{1, \dots, N - 1\}$.
\end{proof}

\subsection{Generalized S/BW canonical circuits $\Fqr$}
\label{app_sub:fqr}

We now turn to the canonical form $\Fqr$. For a given $N \in \mathbb{N}$,
define the set of allowed $(q, r)$ pairs (see Theorem~\ref{th:equivalence_qr})
\begin{align}
	Q_N = \{ (q, r); \ \  & 2 \leq q \leq N, 1 \leq r < q, \nonumber          \\
	                      & \mathrm{gcd}(q, N) = q, \mathrm{gcd}(q, r) = 1\}.
\end{align}
The allowed $(q, r)$ generate either a generalized S circuit
\begin{align}
	Q_N^{(\mathrm{S})} = \{(N, r); 1 \leq r < N, \mathrm{gcd}(N, r) = 1\}
\end{align}
or a generalized BW circuit
\begin{align}
	Q_N^{(\mathrm{BW})} = \{ (q, r); \ \  & 2 \leq q < N, 1 \leq r < q, \nonumber             \\
	                                      & \mathrm{gcd}(q, N) = q, \mathrm{gcd}(q, r) = 1\}.
\end{align}
Which means that $Q_N = Q_N^{(\mathrm{S})} \cup Q_N^{(\mathrm{BW})}$, where
$Q_N^{(\mathrm{S})} \cap Q_N^{(\mathrm{BW})} = \emptyset$.

The first thing to check is that $\Fqr$ is well defined, i.\ e.\ if Floquet operators
$\Fqr$ actually correspond to simple circuits.
\begin{lemma}
	$\Fqr$ for $N \in \mathbb{N}$ and $(q, r) \in Q_N$ are simple circuits,
	i.\ e.\ contain every gate on neighboring sites exactly once. \label{lm:good_def}
\end{lemma}
\begin{proof}
	To show that $\Fqr$ are simple circuits, we thus have to check that the
	sequence~\eqref{eq:qrn_seq} contains every positive integer less than or
	equal to $N$ exactly once.

	Let us first consider the generalized S case, where $q = N$ and
	$\mathrm{gcd}(N, r) = 1$. Because of that, it is intuitively clear, that
	translations by $r$ must be ``ergodic'' and thus generate all gates.
	More formally, this means that the least common multiple
	$\mathrm{lcm}(N, r) = rN/\mathrm{gcd}(N, r) = rN$. Since $F_{N, r}$ is
	generated by translating gate $1$ by $r$ sites, the lowest number of translations after
	which gate $1$ (modulo $N$) is generated again must be $N$ (since the smallest
	solution to $kr = 0 \pmod{N}$ is exactly $k = \mathrm{lcm}(N, r)/r$). Thus,
	before looping back to gate $1$ we generate $N$ gates. By an analogous argument,
	any gate is generated again only after $N$ translations, which means that all the
	generated gates are different.

	Let us now consider the generalized BW case where $q \neq N$. In this case
	$\mathrm{lcm}(q, N) = N$, which means that with translations by $q$ we
	generate $N/q$ different gates. These are exactly gates $1$ plus all the
	multiples of $q$ less than or equal to $N$, so gates $1 + qk$, $k = 0, \dots,
		N/q-1$. By translating them by any $r$, such that $1 \leq r < q$, we
	generate all gates $1 + r + qk$, which since $r < q$ cannot be equal to any
	gates generated with $r = 0$. More formally, this is
	true because $1 + r + qk \not\equiv 1 + qk \pmod{q}$ necessarily implies $1 + r + qk \neq 1 + qk$. In
	order to generate all gates, $1 + kr$ for $k \in \{0, 1, 2, \dots, q - 1\}$
	must now take all the possible values modulo $q$. Analogous to the
	generalized S case, this indeed does happen when $\mathrm{gcd}(q, r) = 1$.
\end{proof}

Since Theorem~\ref{th:equivalence_n} implies that there are $(N - 1)$ non-equivalent
simple circuits, a required condition for $\Fqr$ to be a canonical form is that
there are $(N - 1)$ allowed $(q, r)$.
\begin{lemma}
	$|Q_N| = N - 1$. \label{lm:q_size}
\end{lemma}
\begin{proof}
	Let $\varphi(N) = |\{k; 1 \leq k < N, \mathrm{gcd}(N, k) = 1\}|$ denote
	the number of natural numbers less than $N$ and coprime with $N$, also called
	the Euler's totient or Euler's phi function.

	We can now express the number of generalized staircase circuit with Euler's phi function
	\begin{align}
		\left|Q_N^{(\mathrm{S})}\right| = |\{(N, r); 1 \leq r < N, \mathrm{gcd}(N, r) = 1\}| = \varphi(N)
	\end{align}
	and also the number of generalized BW circuits
	\begin{align}
		\left|Q_N^{(\mathrm{BW})}\right| = & \ |\{ (q, r); \ \   2 \leq q < N, 1 \leq r < q,                                \nonumber    \\
		                                   & \quad \mathrm{gcd}(q, N) = q, \mathrm{gcd}(q, r) = 1\}|                         \nonumber   \\
		=                                  & \ \sum_{q = 2, q | N}^{N - 1} |\{(q, r); 1 \leq r < q, \mathrm{gcd}(q, r) = 1\}|  \nonumber \\
		=                                  & \ \sum_{q = 2, q | N}^{N - 1} \varphi(q)
		% & |\{(q, r) \in \mathbb{N} \times \mathbb{N} ; 1 < q < N, q | N, r < q, \mathrm{gcd}(q, r) = 1\}| = \nonumber \\
		% & =\sum_{q | N, q \neq N, q \neq 1} \varphi(q) = \sum_{q | N, q \neq N} \varphi(q) - \varphi(1) =             \\
		% & = \sum_{q | N, q \neq N} \varphi(q) - 1. \nonumber
	\end{align}
	where $q | N$ denotes that $q$ divides $N$ (equivalently $\mathrm{gcd}(q, N) = q$).

	Thus, since $Q_N^{(\mathrm{S})} \cap Q_N^{(\mathrm{BW})} = \emptyset$
	\begin{align}
		|Q_N| & = \left|Q_N^{(\mathrm{S})}\right| + \left|Q_N^{(\mathrm{BW})}\right| = \sum_{q = 2, q | N}^{N} \varphi(q) \nonumber \\
		      & = \sum_{q|N} \varphi(q) - \varphi(1) = N - 1,
	\end{align}
	where we used the well-known theorem that $\sum_{q | N} \varphi(q) = N$~\cite{rosenElementaryNumberTheory2011} and
	$\varphi(1) = 1$.
\end{proof}

We now wish to show that the invariant $C(\Fqr)$ is different for all allowed
$(q, r)$, which is the last important statement required for the proof of
Theorem~\ref{th:equivalence_qr}. We do this in two steps, first for the
generalized S in Lemma~\ref{lm:c_for_s} and then for generalized BW in
Lemma~\ref{lm:c_for_bw}, finally combining both in
Lemma~\ref{lm:c_for_qr}.

\begin{lemma}
	In the generalized S case, $(q, r) \in Q_N^{(\mathrm{S})}$,
	\begin{align}
		C(F_{q, r}) & = 1 + |\{\text{gates appearing after gate $N$}\}|                \nonumber \\
		            & \in \{p; p < N, \mathrm{gcd}(p, N) = 1\} \label{eq:gen_s_c_set},
	\end{align}
	and is different for different $r$.
	Here all the values in the set~\eqref{eq:gen_s_c_set} are taken for some allowed $r$.
	\label{lm:c_for_s}
\end{lemma}
\begin{proof}
	Since we are considering the generalized S case: $q = N$, $r < N$ and
	$\mathrm{gcd}(N, r) = 1$.

	We first want to consider if we know that a certain gate is a member of
	$\tilde C(F_{N, r})$, what can we say about the membership of its time successor
	and predecessor. Let $F_{N, r} = (i_1, \dots, i_N)$. Let $i_l \in \tilde
		C(F_{N, r})$, which means that its right neighbor appears in $F_{N, r}$
	before it, $\exists i_k = i_l + 1, k < l$ (gate numbers are taken modulo
	$N$, but indices are not). Here $l \neq 1$, since in that case, such $k < l = 1$ never exists.
	Let us now consider the membership of $i_l$'s successor and predecessor:
	\begin{enumerate}[(a)]
		\item If $i_{l + 1} = i_l + r$ is a valid gate (i.\ e.\ its index is valid,
		      which means $l < N$), then $i_{k + 1} = i_{k} + r = i_l + 1 + r = i_{l +
					      1} + 1$, is the right neighbor of $i_{l + 1}$ and $k + 1 < l + 1$, which
		      implies $i_{l + 1} \in \tilde C(F_{N, r})$. In other words:
		      \textit{If $i_l$'s time successor exists (i.\ e.\ $i_l$ is not
			      the last gate in $F$), it is also a member of $\tilde
				      C(F_{N, r})$}.
		\item $i_{l - 1} = i_l - r$, then if $i_{k - 1} = i_{k} - r = i_l + 1 - r =
			      i_{l - 1} + 1$ is a valid gate (i.\ e.\ $k > 1$), it is the right neighbor
		      of $i_{l - 1}$ and $k - 1 < l - 1$, so clearly $i_{l - 1} \in \tilde
			      C(F_{N, r})$. In other words: \textit{If $i_l$'s right neighbor is not $i_1$, then
			      it's time predecessor is also a member of $\tilde C(F_{N, r})$}.
	\end{enumerate}
	In case (b) $i_{k - 1}$ is not valid only if $k = 1$. In the case of $\Fqr$,
	$i_1 = 1$ and thus $i_l = N$. Since always $N \in \tilde C(F_{N, r})$ (its
	right neighbor is $1 = i_1$), by iterating (a), all gates appearing after
	gate $N$ are also members of $\tilde C(F_{N, r})$. If any gate appearing
	before gate $N$ would be a member of $\tilde C(F_{N, r})$, iterating (b)
	would eventually lead to $i_1 = 1 \in \tilde C(i_1, \dots, i_N)$, which
	cannot happen, thus leading to a contradiction. We have shown:
	\begin{align}
		C(F_{N, r}) & = 1 + |\{\text{gates appearing after gate $N$}\}| = \nonumber \\
		            & =N - k_N + 1,
	\end{align}
	where $k_N$ is the index of gate $N$, $i_{k_N} = N$. We have thus shown the first part of
	the lemma.

	We now want to determine what are the possible values of $C(F_{N, r})$.
	In order to do that, we must only find the possible values of $k_N$.
	By definition, we get gate $N$ after $k_N - 1$ translations of gate $1$ by $r$
	\begin{align}
		1 + (k_N - 1)r        & \equiv 0 \pmod{N}, \nonumber \\
		\implies -(k_N - 1) r & \equiv 1 \pmod{N}.
	\end{align}
	Thus, $-r$ is a modular multiplicative inverse of $k_N - 1$.
	According to a well known theorem~\cite{rosenElementaryNumberTheory2011},
	a requirement for modular multiplicative inverses to exists is $\mathrm{gcd}(k_N - 1, N) = 1$. Therefore
	\begin{align}
		\mathrm{gcd}(C(F_{N,r}), N) & = \mathrm{gcd}(N - k_N + 1, N) = \nonumber \\
		                            & =\mathrm{gcd}(k_N - 1, N) = 1.
	\end{align}
	In other words, $C(F_{N, r})$ is coprime with $N$.
	According to Lemma~\ref{lm:c_values}, $C(F) \leq N - 1$, which means that $C(F_{N, r})$
	can only be a number less than $N$ and coprime with $N$
	\begin{align}
		C(F_{N, r}) \in \{p; p < N, \mathrm{gcd}(p, N) = 1\}. \label{eq:c_vals_r0}
	\end{align}

	Let us now show that $k_N$ is different for different $F_{N, r}$ and $F_{N, \tilde r}$ by
	contradiction. Let us suppose otherwise, this means that $k_N - 1$ translations by $r$ and
	by $\tilde r$ must generate the same gate number modulo $N$.
	\begin{align}
		(k_N - 1) r \equiv (k_N - 1) \tilde r \pmod{N}.
	\end{align}
	Since $\mathrm{gcd}(k_N - 1, N) = 1$, we can divide the equation by $k_N - 1$~\cite{rosenElementaryNumberTheory2011}
	\begin{align}
		r \equiv \tilde r \pmod{N}.
	\end{align}
	By definition of allowed $F_{N, r}$ circuits, $r, \tilde r < N$, which means
	that $r = \tilde r$, leading to a contradiction.

	We have thus shown that for different allowed $r$,
	$C(F_{N, r})$ are different. Moreover, in Eq.~\eqref{eq:c_vals_r0} we have
	shown that the value of $C(F_{N, r})$ is a member of a set with exactly
	the same number of elements as allowed $r$ (see $\left| Q_N^{(\mathrm{S})} \right|$ in
	the proof of Lemma~\ref{lm:q_size}). This means that
	$C(F_{N, r})$ takes all the values from the set in Eq.~\eqref{eq:c_vals_r0} if
	we let $r$ be all the allowed $r$ values.
\end{proof}

\begin{lemma}
	In the generalized BW case, $(q, r) \in Q_N^{(\mathrm{BW})}$, for fixed $q$
	\begin{align}
		C(F_{q, r}) & \in \left\{\frac{N}{q}p; p < q, \mathrm{gcd}(p, q) = 1\right\}. \label{eq:gen_bw_c_set}
	\end{align}
	The values are different for different $r$ and are all taken for some allowed $r$.
	\label{lm:c_for_bw}
\end{lemma}
\begin{proof}
	Since we are considering the generalized BW case: $q | N, r < q,
		\mathrm{gcd}(q, r) = 1$.

	In this case, we can divide the circuit in $q$ blocks (layers) of $N/q$ time
	consecutive gates. The first block is generated by translating gate $1$ by
	$q$ (modulo $N$) until we reach gate $1$ again. The second block consists of
	all the gates from the first block translated by $r$. This means that the
	first block can be mapped to the subgroup of the cyclic group $\mathbb{Z}_N
		= \left\{0, \dots, N -1\right\}$ of order $N/q$, where group addition is taken
	modulo $N$. The other blocks are cosets of this subgroup (since according to
	Lemma~\ref{lm:good_def}, they must be disjoint), implying that the first $q$
	gate numbers are contained in different blocks. We can therefore denote the
	blocks with the minimal gate number contained in it (and a tilde for clarity)
	$\tilde 1, \tilde 2, \dots, \tilde q$.

	We now wish to consider a similar thing as in the proof of
	Lemma~\ref{lm:c_for_s}, but in the context of a block, i.\ e.\ given that a
	gate is a member of $\tilde C$, what can we say about the membership of its
	time successor and predecessor. We will see that the main difference is that
	this time, the notion time successor/predecessor can be taken \textit{inside
		a given block}, so with indices modulo $N/q$.

	Let us suppose that a gate with index $l$ (this index is now taken in
	the context of a block, meaning modulo $N/q$) in block $\tilde l$ (modulo
	$q$) is a member of $\tilde C(\Fqr)$, $i_l^{\left(\tilde l\right)} \in \tilde
		C(\Fqr)$. This means that its right neighbor must appear before it, $\exists
		i_{k}^{\left(\tilde k\right)} = i_l^{\left(\tilde l\right)} + 1, \tilde k <
		\tilde l$ (if such $k$ would exists in block $\tilde k = \tilde l$, the block
	would contain all $N$ gate numbers, which is not possible in the generalized BW
	case). Analogously to the generalized S case, $i_{l +
				1}^{\left(\tilde l\right)} = i_l^{\left(\tilde l\right)} + q$ and $i_{k +
				1}^{\left(\tilde k\right)} = i_k^{\left(\tilde k\right)} + q =
		i_l^{\left(\tilde l\right)} + q + 1 = i_{l + 1}^{\left(\tilde l\right)} +
		1$ and thus $i_{l + 1}^{\left(\tilde l\right)} \in \tilde C(\Fqr)$.
	Crucially, indices $k$ and $l$ here are taken modulo $N/q$ because translating
	the last gate of the block by $q$ (modulo $N$) yields the first element
	of the block. Iterating this, we can conclude that if a gate from a
	block is contained in $\tilde C(\Fqr)$, then all the gates from this block are
	contained in $\tilde C(\Fqr)$.

	We have therefore shown that by relabelling the considered circuit with
	$\tilde 1, \tilde 2, \dots, \tilde q$, the considered circuit behaves exactly
	the same as is in the generalized S case. Using Lemma~\ref{lm:c_for_s}, for a
	fixed
	$q$ and $N$
	\begin{align}
		C(\Fqr) \in \left\{\frac{N}{q} p; p < q, \mathrm{gcd}(p, q) = 1\right\}, \label{eq:possible_c_r}
	\end{align}
	where we took into the account that the number of gates in a block is $N/q$.
	Also according to Lemma~\ref{lm:c_for_s}, all the values from the set in
	Eq.~\eqref{eq:possible_c_r} are taken for some $r$ and are different for
	different $r$.
\end{proof}

\begin{lemma}
	$C(\Fqr)$ is different for all $(q, r) \in Q_N$ for a given $N$.
	\label{lm:c_for_qr}
\end{lemma}
\begin{proof}
	We have calculated all the allowed values of $C(\Fqr)$ and shown that they
	are different for different $r$ and a fixed $q$ in Lemma~\ref{lm:c_for_s} and Lemma~\ref{lm:c_for_bw}.
	The only thing left to do is to show that for different $q$, the sets \eqref{eq:gen_s_c_set}
	and \eqref{eq:gen_bw_c_set} are disjoint.

	From Eq.~\eqref{eq:gen_bw_c_set}, we can see that in the generalized BW case
	$\mathrm{gcd}(C(\Fqr), N) \geq N/q$, meaning that all the values of $C(\Fqr)$
	must be different than in the generalized S case ($q = N$) where $\mathrm{gcd}(C(F_{N,
				r}), N) = 1$ (Eq.~\eqref{eq:gen_s_c_set}).

	The only thing left is to check that the sets in the BW case
	(Eq.~\eqref{eq:gen_bw_c_set}) are disjoint for two different $q_1, q_2 | N$.
	We can do this by finding a contradiction in the converse case. Let us suppose
	that we would find the same value of $C$ for different $q_1, q_2$, this means
	that for some $p_1, p_2, \mathrm{gcd}(p_1, q_1) = \mathrm{gcd}(p_2, q_2) = 1$
	we must have
	\begin{align}
		\frac{N}{q_1} p_1 & = \frac{N}{q_2}p_2,         \\
		\implies q_2 p_1  & = q_1 p_2. \label{eq:pq_eq}
	\end{align}
	This implies that $p_1 | q_1 p_2$, but since $\mathrm{gcd}(p_1, q_1) = 1$, we
	see that $p_1 | p_2$. Symmetrically, we can show $p_2 | p_1$, which then
	implies $p_1 = p_2$. From Eq.~\eqref{eq:pq_eq} we finally get $q_1 = q_2$, a
	contradiction. Therefore, $C(\Fqr)$ must also be different for different $q$
	in the BW case.
\end{proof}

Finally, we can put all the previous lemmas together and prove the main theorem.
	{
		\renewcommand*{\thetheorem}{\ref{th:equivalence_qr}}
		\begin{theorem}[restated from the main text]
			Any simple circuit on $N$ sites with periodic boundary conditions is
			equivalent to exactly one of the $N - 1$ canonical simple circuits
			having Floquet propagator $\Fqr$, where $(q, r) \in Q_N$.
		\end{theorem}
		\addtocounter{theorem}{-1}
	}
\begin{proof}
	The fact that $\Fqr$ are simple circuits follows from Lemma~\ref{lm:good_def} and the number of them
	follows from Lemma~\ref{lm:q_size}.

	According to Lemma~\ref{lm:c_for_qr}, $C(\Fqr)$ is different for all allowed
	$(q, r)$ and a given $N$. Since according to Lemma~\ref{lm:c_conserv} $C$ is
	conserved under equivalence transformations and, according to
	Eq.~\eqref{eq:c_fp_val}, is different for all $F_p$, different $\Fqr$
	are equivalent to different $F_p$ (according to
	Theorem~\ref{th:equivalence_n}, they must be equivalent to some $F_p$).
	Moreover, since there are $(N - 1)$ $\Fqr$ and $(N - 1)$ $F_p$, every
	$F_p$ must be equivalent to some $\Fqr$, i.\ e.\ there exists a bijective equivalence
	mapping between $F_{q, r}$ and $F_p$.

	Given any simple circuit $F$, according to Theorem~\ref{th:equivalence_n}, we know $F \cong F_p$
	for some $p$. We have shown that $F_p \cong \Fqr$ for some $q, r$. Therefore, $F \cong \Fqr$ for
	some $q, r$.
\end{proof}

In contrast to the proof of Theorem~\ref{th:equivalence_n}, the proof of
Theorem~\ref{th:equivalence_qr} is not constructive, i.\ e.\ it cannot be used
as an algorithm to transform a given $F$ to the equivalent $\Fqr$. The most
convenient way to do this in practice is via the invariant $C(F)$ defined in
Eq.~\eqref{eq:c_def}, which can be numerically calculated in linear time in
$N$. One can then calculate $C(F)$ and $C(\Fqr)$ for the all $(N - 1)$ allowed
$(q, r)$ in quadratic time. Finally, $F$ must be equivalent to the $\Fqr$ with
the same value of $C$. As an example, the values of $C(\Fqr)$ are included in Fig.~\ref{fig:fqr_diagram_10}.

\end{document}